\documentclass[fleqn,usenatbib]{mnras}

\usepackage{newtxtext,newtxmath}
\usepackage{xcolor}
\usepackage{bm}
\usepackage{color}
\usepackage{graphicx}

\newcommand{\ie}{i.e.,~}
\usepackage{soul}
\usepackage[T1]{fontenc}

\DeclareRobustCommand{\VAN}[3]{#2}
\let\VANthebibliography\thebibliography
\def\thebibliography{\DeclareRobustCommand{\VAN}[3]{##3}\VANthebibliography}

\usepackage{amsmath}	
\usepackage[toc,page]{appendix}

\title[Relativistic Common-envelope]{Relativistic Common-envelope Dynamics of a Stellar-mass Black Hole II: Kerr Black Hole}
\author[Cruz-Osorio et al.]{Alejandro Cruz-Osorio,$^{1}$\thanks{E-mail: aosorio@astro.unam.mx}
F. D. Lora-Clavijo $^{2}$\thanks{fadulora@uis.edu.co}\\
$^{1}$ Instituto de Astronom\'ia, Universidad Nacional Aut\'onoma de M\'exico, AP 70-264, Ciudad de M\'exico 04510, Mexico \\
$^{2}$ Grupo de Investigaci\'on en Relatividad y Gravitaci\'on, Escuela de F\'isica, \\
Universidad Industrial de Santander A. A. 678, Bucaramanga 680002, Colombia \\
}

\begin{document}
\maketitle

\begin{abstract}
The common envelope phase plays a critical role in binary system evolution. 
In this study, we investigate the mass and momentum accretion rates during the interaction between a stellar-mass black hole and the envelope of a red supergiant using simplified two-dimensional simulations. We explore various Mach numbers and density gradients, finding that our simulations align with previous Bondi–Hoyle–Lyttleton accretion analyses. We observe the formation of a shock cone in the downstream flow, bow shocks in specific configurations, and subsonic regions within shocked flows. The shock cone is dragged when significant pressure gradients are present in the common envelope, with additional dragging near the black hole for highly rotating cases. We provide analytical fits for mass and momentum accretion rates, as well as bremsstrahlung luminosity, as functions of black hole spin, density gradients, and Mach number, offering first insights into general relativistic hydrodynamics modelling of the secular evolution of the common-envelope phase.
\end{abstract}

\section{Introduction}
\label{sec:introduction}
Common envelope (CE) evolution is believed to be an important stage in 
the evolution of binary/multiple stellar systems. This almost certainly includes the 
progenitors of Type Ia supernovae, X-ray binaries and double neutron stars. 
The phenomenon consists in a dramatic event that can occur in close binary star systems. 
When one of the stars in the system, typically a giant star that has 
reached the end of its main sequence life, expands dramatically, 
it can engulf its companion star or compact object in a 
giant envelope of gas \citep{Paczynski76}. 
Understanding common envelope phases is crucial for explaining 
the observed properties of various binary systems and for modeling 
the evolutionary pathways of compact object binaries. 
The dynamics during these phases are complex and depend on 
factors such as mass transfer rates, stellar masses, and orbital parameters \citep{Ivanova2013}.

The study of common-envelope dynamics delves into the complex 
interplay of gravitational forces, relativistic effects, and hydrodynamic 
interactions that occur when a compact object, such as a neutron star or 
black hole, becomes engulfed within the extended envelope of a massive 
companion star. This phenomenon is a critical stage in the evolution of 
binary star systems, particularly in scenarios that lead to the formation 
of gravitational wave sources, short gamma-ray bursts, and compact 
object mergers \citep{MacLeod2015, MacLeod2015b, MacLeod2017, Murguia-Berthier2017845}.

The spin of black holes (or other compact objects) is a critical factor 
influencing the evolution of a common envelope during binary stellar 
interactions \citep{Bavera2020}. The black hole’s spin could also 
interact with the gas envelope via tidal forces, redistributing angular 
momentum within the system \citep{Gafton2019}. Additionally, 
during the CE phase, accretion onto the black hole can power 
relativistic jets with spin directly influencing jet efficiency \citep{Kaaz2022} 
\citep[see also ][]{LopezCamara2020}. 
These jets inject energy and momentum into the envelope, facilitating 
ejection or modifying its structure. Understanding these dynamics is 
key to explaining the diversity of observed compact-object binaries 
and their gravitational wave signals.

The relativistic Bondi-Hoyle-Lyttleton (BHL) accretion is a key concept 
in understanding the evolution of the CE phase, particularly when one 
of the objects involved is a neutron star or black hole. The BHL accretion 
formalism describes the rate at which the compact object accretes matter 
from the surrounding envelope. The accretion rate depends on the local 
density, velocity, and sound speed of the material \citep{Bondi52_new,Bondi1944,Hoyle1939,Edgar2004,Foglizzo2005,Foglizzo1999}.

The relativistic approach to BHL accretion has proven essential for studying 
accretion in regions with strong gravitational fields, where Newtonian methods 
are insufficient. Early foundational work by \cite{Petrich1988, Petrich89} 
explored the accretion patterns of relativistic gas flowing onto a black hole. 
These efforts were later refined by \cite{Font98a, Font98c, Font98d, Font1999b}, 
who employed more accurate methods and extended the analysis to include 
both axial and equatorial symmetries. In a relativistic astrophysical context,
\cite{Donmez2010} linked shock-cone vibrations to quasi-periodic 
oscillations (QPOs) and identified a "flip-flop" type of unstable oscillation 
in the shock cone. However, \cite{Cruz2012} demonstrated that the flip-flop 
oscillation is coordinate-dependent, disappearing when Kerr–Schild 
coordinates are used to describe the rotating black hole. Subsequent 
studies have further examined shock-cone oscillations as potential 
sources of both low- and high-frequency QPOs \citep{Lora-Clavijo2013}, 
while other efforts have incorporated astrophysical factors such as 
magnetic fields \citep{Penner2011}, radiative effects \citep{Zanotti2011}, 
full general relativity in scenarios like supermassive black hole binary 
mergers \citep{Farris2010}, and density and velocity gradients 
\citep{Lora-Clavijo2015,Cruz2016}. Additional efforts have analyzed the 
influence of small rigid bodies near the black hole \citep{Cruz2017}, 
explored BHL accretion during the common-envelope phase of binary 
system evolution \citep{Cruz2020b}, offering insights into its role in the 
formation and evolution of compact objects, and investigated the impact 
of a scalar field in the BHL accretion onto a rotating black hole \citep{Cruz2023}.

The inclusion of magnetized media in BHL accretion has further advanced 
the understanding of relativistic accretion dynamics. Studies focusing on 
general-relativistic BHL accretion in a toroidally magnetized medium 
have shown how magnetic tension and pressure influence shock-cone 
structures and energy transport \citep{Kim2024}. Moreover, three-dimensional 
general-relativistic magnetohydrodynamic (GRMHD) simulations have 
provided significant insights into jet formation during BHL accretion \citep{Kaaz2022}. 
These simulations demonstrate that accretion onto rotating black holes 
in magnetized environments can naturally produce relativistic jets, 
shedding light on the mechanisms driving jet formation in systems 
such as active galactic nuclei and microquasars. Together, these 
developments emphasize the growing importance of magnetohydrodynamic 
effects and advanced numerical techniques in understanding relativistic accretion processes.

In this work, we extend the study of \cite{Cruz2020b}, by quantifying 
the accretion dynamics and radiative output of a stellar-mass rotating black 
hole embedded in the common envelope of a red supergiant. We systematically 
explore the dependence of the black hole spin $a_{\star}$, Mach number $\mathcal{M}$, 
and density gradient $\varepsilon$. 
We measure the mass and momentum accretion rates, the shock cone morphology, 
the presence and location of bow shocks, and the bremsstrahlung luminosity. 
The explored parameter space spans sonic to supersonic relative motions and 
density gradients representative of red supergiant envelopes. In this work, we adopt a 
two-dimensional geometry as a first step toward more realistic three-dimensional 
scenarios. Despite this limitation, our idealized model constitutes the first numerical 
study of this problem in the general relativistic regime. The direct application of our 
results to population synthesis models will require calibration with three-dimensional 
simulations, which we plan to investigate in future work.\\

The structure of this paper is as follows: Section \ref{sec:GRHD} introduces 
the relativistic hydrodynamics equations and outlines the key features of the 
common-envelope scenario. It further describes the construction of the initial 
data, the boundary conditions implemented, the computational infrastructure 
utilized, and the numerical methods, including the grid configuration and 
resolution strategies adopted to ensure accuracy and stability. In Section 
\ref{sec:Results}, we present our results, focusing on the fluid dynamics 
morphology, the accretion rates, and other relevant quantities, such as the 
deviation angle of the shock cone and the resulting bremsstrahlung luminosity. 
These quantities are analyzed as functions of the dimensionless spin parameter 
$a_{\star}$, the Mach number, and the density gradient parameter. 
Finally, we present our main concluding remarks in Section \ref{sec:conclusion}. 

Unless stated otherwise we use geometrised units in which the light speed, 
Newton's constant, and the mass of the black hole are equal to one, $c=G=M=1$, 
the Kerr metric has the signature $(-,+,+,+)$.

\section{General relativistic hydrodynamics simulations}
\label{sec:GRHD}

The numerical simulations are performed by solving the general relativistic 
hydrodynamics equations, following the methodology established in our 
previous studies \citep{Cruz2020b,Cruz2023}. In a generic curved spacetime, 
the covariant fundamental equations can be expressed as conservation laws 
for the energy-momentum tensor, $T^{\mu \nu}$, and the rest-mass current, 
$J^{\mu} = \rho u^{\mu} $ \citep{Rezzolla_book:2013},
\begin{eqnarray}
  \nabla_{\mu}(T^{\mu\nu}) &=& 0\,,\label{eq:SEcons} \\
  \nabla_{\mu}(J^{\mu})  &=& 0\, ,
  \label{eq:masscons}
\end{eqnarray}
being $\rho$ and $u^{\mu}$  the rest mass density and the four-velocity components of the fluid.  
The energy-momentum tensor for a perfect fluid, neglecting shear and viscosity effects, is given by:
\begin{eqnarray}
  T^{\mu \nu} =\rho h u^{\mu}u^{\nu} + pg^{\mu\nu}\, .
  \label{eq:stress-energy}
\end{eqnarray}
Here $h=1+\epsilon+p/\rho$ represents the specific enthalpy, 
where $\epsilon$ is the specific internal energy,  
$p$ is the pressure, and $g^{\mu \nu}$ are the components of the 
spacetime metric tensor.  For this work, we assume that the state 
variables of the system satisfy the equation of state for an ideal gas, 
expressed as: $p = \rho \epsilon (\gamma -1)$, where $\gamma$ is 
the adiabatic index of the gas. Specifically, we will use $\Gamma= 5/3$, 
which corresponds to a cold degenerate electron fluid \citep{Rezzolla_book:2013}.

In this work, we adopt the 3+1 spacetime decomposition to express the evolution 
equations in a conservative form, following the Valencia 
formulation \citep{Banyuls97}. We consider a Kerr black hole characterized 
by a dimensionless spin parameter, $a_{\star}$, and mass $M$. 

The matter is assumed to behave as a test fluid and therefore has a 
negligible effect on the background spacetime geometry. To justify this 
approximation, we compute the compactness of the red supergiant primary. 
For $M_{\rm star} = 16M_{\odot}$ and $R_{\rm star} = 800R_{\odot}$, the 
stellar compactness is 
$\mathcal{C}_{\rm RSG} = GM_{\rm star}/(R_{\rm star}c^2) \approx 4.3 \times 10^{-8}$. 
In comparison, the stellar-mass black hole with $M = 4M_{\odot}$ has a 
compactness of $\mathcal{C}_{\rm BH} = 0.5$ at its Schwarzschild radius. 
The ratio $\mathcal{C}_{\rm RSG}/\mathcal{C}_{\rm BH} \approx 8.5 \times 10^{-8} \ll 1$ 
indicates that the black hole curvature dominates the local spacetime 
geometry. Therefore, the gravitational influence of the common-envelope 
material on the background spacetime is negligible, and the test-fluid 
approximation is well justified.

Following \citep{Cruz2020b, Cruz2023}, we model the rest-mass density distribution of the common 
envelope around the rotating black hole with a constant-density core that decreases exponentially 
toward the stellar surface. In a coordinate system centered at the star's center, the rest-mass density is given by:
\begin{equation}
\label{eq:dence}
\rho(r)=
\begin{cases}
  \bar{\rho}={\rm const.} \quad &{\rm for}\quad 0 < r < \bar{r}  \\
  \rho_{0} \exp{\left[-\epsilon_{\rho}(r-r_{0})/r_{\rm acc} \right]}\,.
  \quad &{\rm for}\quad \bar{r} \leq r \leq R_{\rm star}
  \end{cases}
\end{equation}
Here, $r_0 > \bar{r}$ is the radial position of the black hole within the envelope, 
$\rho_{0} = \rho(r=r_0)$ is the rest-mass density at that position, and $\bar{\rho} = \rho(r=\bar{r}) > \rho_0$ 
is the density of the constant-density core.The characteristic length scale over which the 
black hole's gravity dominates the surrounding fluid dynamics is the accretion radius, defined as:
$$r_{\rm acc} = \frac{GM}{c_{s,\infty}^2 + v_{\infty}^2}$$
where $c_{s,\infty}$ is the asymptotic sound speed and $v_{\infty}$ is the asymptotic 
velocity \citep{Hoyle1939, Bondi1944, Cruz2012, Lora-Clavijo2013, Lora-Clavijo2015, Cruz2016}.

The exponential density profile adopted in the outer envelope is motivated 
as a local first-order approximation to any smooth stellar density distribution over
 the characteristic accretion scale, $r_{\rm acc}$. This prescription follows established 
 studies of common envelope accretion \citep{MacLeod2015}  and provides a 
 framework for investigating the impact of density gradient effects through the 
 dimensionless parameter $\epsilon_{\rho}$.

To perform the 2D simulations, we map the spherical polar 
coordinate system $(r,\phi)$ to a cartesian system $(x,y)$, where 
the stellar center is located at $(x_0, y_0)$ \citep[see][for more details on the mapping]{Cruz2012}. 
The common envelope is assumed to move in the positive $x$-direction, 
such that $v^{x}=v_{\infty}$. Additionally, the matter is assumed to 
have a rest-mass gradient only in the $y$-direction (see 
Equation \eqref{eq:dence}). As a result, the initial rest-mass 
density profile is defined as: 
\begin {equation}
\rho_{\rm in} := \rho(y) = \rho_{0} \exp{\left[
-\epsilon_{\rho} y/r_{\rm acc} \right]}, 
\end{equation}
where $y\in [-\bar{r},\bar{r}]$. The initial pressure is determined 
using the definition of the sound speed and the polytropic equation of state:
\begin {equation}
p_{\rm in}=c^{2}_{s, \infty} \frac{\Gamma -1}{\Gamma(\Gamma -1)
-c^{2}_{s, \infty} \Gamma}\,\rho_{\rm in},
\end{equation}
as described in \citep{Cruz2012,Lora-Clavijo2013,Cruz2020b}.

The numerical grid is constructed using radial and azimuthal 
coordinates and is uniformly spaced with $N_r \times N_{\phi} = 4080 \times 256$ cells. 
The radial grid extends from $r_{\rm exc}$ to $r_{\rm max}$, where $r_{\rm exc}$ is 
the excision radius, located inside the event horizon of the black hole. 
Specifically, $r_{\rm exc}= 0.95 r_{\rm EH}$, with $r_{\rm EH}$ being the 
black hole's event horizon. The outer boundary of the radial grid is set at 
$r_{\rm max}=10\, r_{\rm acc}$, where  $r_{\rm  acc}$ is the accretion 
radius (see Tables \ref{tab:params1}  and \ref{tab:params2}). The angular 
coordinate, $\phi$, spans the full range $\phi \in [0 , 2 \pi]$. The effective 
resolution of the grid is $(\Delta r, \Delta \phi)= (0.1 \, M, 0.0386 \, {\rm rad})$ in 
geometrized units, or, equivalently, $\Delta r \sim 0.6 \, {\rm km}$.   
The timestep size $\Delta t$, is constant and satisfies the Courant-Friedrichs-Lewy (CFL) 
stability condition, defined as $\Delta t = \tfrac{1}{4}\, \min(\Delta r, \Delta \phi)$. 
For details on the consistency and convergence tests for the common envelope setup, see \citep{Cruz2020b}.

The general relativistic hydrodynamics (GRHD) equations \eqref{eq:SEcons}, 
expressed in conservative form with a 3+1 spacetime split, are solved using 
the \texttt{Black Hole Accretion Code} (\texttt{BHAC}) \citep{Porth2017, Olivares2019}. 
This code is a multidimensional extension of the \texttt{MPI-AMRVAC} 
framework \citep{Porth2014, Keppens2012}, designed to evolve the 
GRMHD equations in a finite volume representation. 
For the numerical fluxes, we employ the Harten, Lax, van Leer, 
and Einfeldt (HLLE) approximate Riemann solver \citep{Harten83, Einfeldt1991}, 
coupled with the "minmod" second-order total variation 
diminishing (TVD) reconstruction at cell interfaces. Time evolution of 
the partial differential equations is performed using the method-of-lines 
approach in combination with a fourth-order Runge-Kutta scheme, 
ensuring the TVD property \citep{Shu88}.

\begin{table}
  \begin{center}
    \caption{This study presents a summary of 30 numerical models, exploring five spin parameters, 
    $a_{\star}$, of the Kerr black hole and and six density gradient values $\epsilon_{\rho}$. 
    The sound speed is fixed at $c_{s, \infty}=0.1$, with an asymptotic Mach number
     ${\cal M}_{\infty}=2$, corresponding to an asymptotic velocity $v_{\infty}=0.2$.
    This choice determines the accretion radius $r_{\rm acc}$, both in mass units and kilometers,
    as reported in the third column of the table. }
\begin{tabular}{lcccccc}
\hline\hline
 Model  & $\epsilon_{\rho}$ &  $r_{\rm acc}  [M \, ( {\rm km}) ]$&$a_{\star}$\\
\hline
\hline
$\texttt{RCE.0.00.5o3.a-15o16}$   &  $0.00$  & $20 \, (119)$ & $-0.9375$  \\
$\texttt{RCE.0.25.5o3.a-15o16}$   &  $0.25$  & $20 \, (119)$ & $-0.9375$\\
$\texttt{RCE.0.50.5o3.a-15o16}$   &  $0.50$  & $20 \, (119)$ & $-0.9375$\\
$\texttt{RCE.0.75.5o3.a-15o16}$   &  $0.75$  & $20 \, (119)$ & $-0.9375$\\
$\texttt{RCE.1.00.5o3.a-15o16}$   &  $1.00$  & $20 \, (119)$ & $-0.9375$\\
$\texttt{RCE.1.25.5o3.a-15o16}$   &  $1.25$  & $20 \, (119)$ & $-0.9375$\\
\hline
$\texttt{RCE.0.00.5o3.a-1o2}$   &  $0.00$  &  $20 \, (119)$ & $-0.5000$  \\
$\texttt{RCE.0.25.5o3.a-1o2}$   &  $0.25$  &  $20 \, (119)$ & $-0.5000$\\
$\texttt{RCE.0.50.5o3.a-1o2}$   &  $0.50$  &  $20 \, (119)$ & $-0.5000$\\
$\texttt{RCE.0.75.5o3.a-1o2}$   &  $0.75$  &  $20 \, (119)$ & $-0.5000$\\
$\texttt{RCE.1.00.5o3.a-1o2}$   &  $1.00$  &  $20 \, (119)$ & $-0.5000$\\
$\texttt{RCE.1.25.5o3.a-1o2}$   &  $1.25$  &  $20 \, (119)$ & $-0.5000$\\
\hline
$\texttt{RCE.0.00.5o3.a00}$   &  $0.00$  &  $20 \, (119)$ & $0.0000$  \\
$\texttt{RCE.0.25.5o3.a00}$   &  $0.25$  &  $20 \, (119)$ & $0.0000$\\
$\texttt{RCE.0.50.5o3.a00}$   &  $0.50$  &  $20 \, (119)$ & $0.0000$\\
$\texttt{RCE.0.75.5o3.a00}$   &  $0.75$  &  $20 \, (119)$ & $0.0000$\\
$\texttt{RCE.1.00.5o3.a00}$   &  $1.00$  &  $20 \, (119)$ & $0.0000$\\
$\texttt{RCE.1.25.5o3.a00}$   &  $1.25$  &  $20 \, (119)$ & $0.0000$\\
\hline
$\texttt{RCE.0.00.5o3.a+1o2}$   &  $0.00$  &  $20 \, (119)$ & $+0.5000$ \\
$\texttt{RCE.0.25.5o3.a+1o2}$   &  $0.25$  &  $20 \, (119)$ & $+0.5000$ \\
$\texttt{RCE.0.50.5o3.a+1o2}$   &  $0.50$  &  $20 \, (119)$ & $+0.5000$\\
$\texttt{RCE.0.75.5o3.a+1o2}$   &  $0.75$  &  $20 \, (119)$ & $+0.5000$\\
$\texttt{RCE.1.00.5o3.a+1o2}$   &  $1.00$  &  $20 \, (119)$ & $+0.5000$\\
$\texttt{RCE.1.25.5o3.a+1o2}$   &  $1.25$   & $20 \, (119)$ & $+0.5000$\\
\hline
$\texttt{RCE.0.00.5o3.a+15o16}$   &  $0.00$  &  $20 \, (119)$ & $+0.9375$\\
$\texttt{RCE.0.25.5o3.a+15o16}$   &  $0.25$  &  $20 \, (119)$ & $+0.9375$\\
$\texttt{RCE.0.50.5o3.a+15o16}$   &  $0.50$  &  $20 \, (119)$ & $+0.9375$\\
$\texttt{RCE.0.75.5o3.a+15o16}$   &  $0.75$  &  $20 \, (119)$ & $+0.9375$\\
$\texttt{RCE.1.00.5o3.a+15o16}$   &  $1.00$  &  $20 \, (119)$ & $+0.9375$\\
$\texttt{RCE.1.25.5o3.a+15o16}$   &  $1.25$  & $20 \, (119)$ & $+0.9375$\\
\hline
\end{tabular}
\label{tab:params1}
\end{center}
\end{table} 

\begin{table}
  \begin{center}
    \caption{This summary covers the 36 models considered, all of which 
    involve a Schwarzschild black hole. Each model assumes a fixed sound speed of $c_{s,
        \infty}=0.07$,  while varying the asymptotic Mach number ${\cal M}_{\infty}$ and 
        the relative scale height of the rest-mass density $\epsilon_{\rho}$. 
        This choice determines the accretion radius, $r_{\rm acc}$, both in mass units 
        and kilometers, as reported in the third column of the table. }
\begin{tabular}{lccccccc}
\hline\hline
 Model  & $\epsilon_{\rho}$ & $r_{\rm acc}  [M \, ( {\rm km}) ]$&$a_{\star}$&${\cal M}_{\infty}$\\
\hline
\hline
$\texttt{RCE.0.00.5o3.a00.m10}$   &  $0.00$  &  $102 \, (603)$ & $0$ &$1.0$ \\
$\texttt{RCE.0.25.5o3.a00.m10}$   &  $0.25$  &  $102 \, (603)$ & $0$&$1.0$\\
$\texttt{RCE.0.50.5o3.a00.m10}$   &  $0.50$  &  $102 \, (603)$ & $0$&$1.0$\\
$\texttt{RCE.0.75.5o3.a00.m10}$   &  $0.75$  &  $102 \, (603)$ & $0$&$1.0$\\
$\texttt{RCE.1.00.5o3.a00.m10}$   &  $1.00$  &  $102 \, (603)$ & $0$&$1.0$\\
$\texttt{RCE.1.25.5o3.a00.m10}$   &  $1.25$  &  $102 \, (603)$ & $0$&$1.0$\\
\hline
$\texttt{RCE.0.00.5o3.a00.m12}$   &  $0.00$  &  $84\, (494)$ & $0$ &$1.2$ \\
$\texttt{RCE.0.25.5o3.a00.m12}$   &  $0.25$  &  $84\, (494)$ & $0$&$1.2$\\
$\texttt{RCE.0.50.5o3.a00.m12}$   &  $0.50$  &  $84\, (494)$ & $0$&$1.2$\\
$\texttt{RCE.0.75.5o3.a00.m12}$   &  $0.75$  &  $84\, (494)$ & $0$&$1.2$\\
$\texttt{RCE.1.00.5o3.a00.m12}$   &  $1.00$  &  $84 \, (494)$ & $0$&$1.2$\\
$\texttt{RCE.1.25.5o3.a00.m12}$   &  $1.25$  &  $84 \, (494)$ & $0$&$1.2$\\
\hline
$\texttt{RCE.0.00.5o3.a00.m14}$   &  $0.00$  &  $69 \, (407)$ & $0$ &$1.4$ \\
$\texttt{RCE.0.25.5o3.a00.m14}$   &  $0.25$  &  $69 \, (407)$ & $0$&$1.4$\\
$\texttt{RCE.0.50.5o3.a00.m14}$   &  $0.50$  &  $69 \, (407)$ & $0$&$1.4$\\
$\texttt{RCE.0.75.5o3.a00.m14}$   &  $0.75$  &  $69 \, (407)$ & $0$&$1.4$\\
$\texttt{RCE.1.00.5o3.a00.m14}$   &  $1.00$  &  $69 \, (407)$ & $0$&$1.4$\\
$\texttt{RCE.1.25.5o3.a00.m14}$   &  $1.25$  &  $69 \, (407)$ & $0$&$1.4$\\
\hline
$\texttt{RCE.0.00.5o3.a00.m16}$   &  $0.00$  &  $57 \, (339)$ & $0$&$1.6$ \\
$\texttt{RCE.0.25.5o3.a00.m16}$   &  $0.25$  &  $57 \, (339)$ & $0$&$1.6$\\
$\texttt{RCE.0.50.5o3.a00.m16}$   &  $0.50$  &  $57 \, (339)$ & $0$&$1.6$\\
$\texttt{RCE.0.75.5o3.a00.m16}$   &  $0.75$  &  $57 \, (339)$ & $0$&$1.6$\\
$\texttt{RCE.1.00.5o3.a00.m16}$   &  $1.00$  &  $57 \, (339)$ & $0$&$1.6$\\
$\texttt{RCE.1.25.5o3.a00.m16}$   &  $1.25$  & $57 \, (339)$ & $0$&$1.6$\\
\hline
$\texttt{RCE.0.00.5o3.a00.m18}$   &  $0.00$  &  $48 \, (284)$ & $0$&$1.8$ \\
$\texttt{RCE.0.25.5o3.a00.m18}$   &  $0.25$  &  $48 \, (284)$ & $0$&$1.8$\\
$\texttt{RCE.0.50.5o3.a00.m18}$   &  $0.50$  &  $48 \, (284)$ & $0$&$1.8$\\
$\texttt{RCE.0.75.5o3.a00.m18}$   &  $0.75$  &  $48 \, (284)$ & $0$&$1.8$\\
$\texttt{RCE.1.00.5o3.a00.m18}$   &  $1.00$  &  $48 \, (284)$ & $0$&$1.8$\\
$\texttt{RCE.1.25.5o3.a00.m18}$   &  $1.25$  &  $48 \, (284)$ & $0$&$1.8$\\
\hline
$\texttt{RCE.0.00.5o3.a00.m20}$   &  $0.00$  &  $41 \, (241)$& $0$&$2.0$ \\
$\texttt{RCE.0.25.5o3.a00.m20}$   &  $0.25$  &  $41 \, (241)$& $0$&$2.0$\\
$\texttt{RCE.0.50.5o3.a00.m20}$   &  $0.50$  &  $41 \, (241)$& $0$&$2.0$\\
$\texttt{RCE.0.75.5o3.a00.m20}$   &  $0.75$  &  $41 \, (241)$& $0$&$2.0$\\
$\texttt{RCE.1.00.5o3.a00.m20}$   &  $1.00$  &  $41 \, (241)$& $0$&$2.0$\\
$\texttt{RCE.1.25.5o3.a00.m20}$   &  $1.25$  &  $41 \, (241)$& $0$&$2.0$\\
\hline
\end{tabular}
\label{tab:params2}
\end{center}
\end{table} 

\subsection{Common envelope setup}

Following the astrophysical motivations presented in \citet{MacLeod2015} and 
\citet{Cruz2020b} concerning common envelope properties, we consider a 
binary system consisting of a red supergiant star and a stellar-mass black hole.
This choice is motivated by key physical conditions expected during the 
common-envelope phase: the presence of an extended, low-density convective 
envelope of red supergiant, transonic to supersonic relative motion of the black 
hole through the envelope, a local density scale height comparable to the accretion 
radius, and the binary mass ratio characteristic of a stellar-mass black hole 
embedded within a red supergiant envelope. The red supergiant has a mass of 
$M_{\rm star} = 16\,M_{\odot}$, while the black hole has a mass of $M = 4\, M_{\odot}$, 
yielding a binary mass ratio of $q := M/M_{\rm star} = 0.25$.
The black hole is initially positioned at 
$r_{0} = R_{\rm star}/2 \simeq 400 R_{\odot} \simeq 2.78 \times 10^8\,{\rm km}$. 
At this location, the rest-mass density is assumed to be 
$\rho(r_0) =: \rho_0 = \rho_{\infty} \simeq 9.51 \times 10^{-9}\,{\rm g/cm}^3$. 
Ado ting this density value and an intermediate density gradient parameter 
of $\epsilon_{\rho} = 0.5$, the integration of the rest-mass density 
profile \eqref{eq:dence} yields a total stellar mass of $16\,M_{\odot}$ 
for the red supergiant. This setup provides a robust framework for 
investigating the interaction between the black hole and the 
extended envelope of the red supergiant, with the mass ratio 
and density gradient being critical parameters that influence 
the common envelope dynamics and subsequent binary evolution.
  
In this work, we investigate three fundamental physical properties 
of the relativistic common envelope interaction with a rotating black hole: 
the density gradients in the red supergiant star as the black hole 
approaches the star's center ($\epsilon_{\rho}$), the asymptotic value of the 
Newtonian Mach number ($\mathcal{M}_{\infty} := v_{\infty}/c_{s,\infty}$), 
and the dimensionless black hole spin ($a_{\star}$). 
The goal is to systematically investigate the influence 
of black hole spin, Mach number, and density gradient on the local
accretion dynamics and shock morphology during the common envelope phase.
Tables \ref{tab:params1} and \ref{tab:params2} summarize the critical 
parameters used in our simulations. 
We examine density gradients $\epsilon_{\rho} = 0.0, 0.25, 0.5, 0.75, 1.0, 1.25$, 
which represent the density variation of the star from its outer to inner layers. 
The Mach number, $\mathcal{M}_{\infty}$, takes values $1.0, 1.2, 1.4, 1.6, 1.8, 2.0$,
covering both sonic (transonic $\mathcal{M}_{\infty} = 1$),  and supersonic regimes 
within the star's envelope. 
For the black hole's dimensionless spin, 
we consider $a_{\star} = -15/16, -1/2, 0, +1/2, +15/16$. 
These parameters provide a comprehensive framework 
for exploring the initial conditions and dynamics of the 
models studied in this work.

We recall that the inner boundary, $r_{\rm exc}$ (the excision radius), 
is always located inside the event horizon, where inward boundary conditions are 
imposed. The outer boundary, $r_{\rm max}$, is divided into two regions: an upstream 
region, through which gas enters the computational domain and where inflow boundary 
conditions are applied, and a downstream region, through which gas leaves the domain 
and where outflow (zero-gradient) boundary conditions are imposed. In addition, periodic 
boundary conditions are adopted in the angular direction \citep[for further details, see][]{Cruz2012,Cruz2020b}.\\
It is worth noting that the 2D simulations captures the essential physical processes 
governing accretion onto a rotating black hole, including radial inflow, angular momentum 
transport in the equatorial plane, and frame-dragging effects, which are strongest near 
the equatorial region. Moreover, the primary goal of this work is to identify qualitative 
trends across a broad parameter space ($a_{\star}$, $\mathcal{M}$, and $\varepsilon$), 
rather than to determine precise absolute values. The 2D approach makes such a parameter 
survey computationally feasible, whereas an equivalent study in 3D at comparable resolution
 would be more expensive. Finally, previous studies of Bondi--Hoyle--Lyttleton accretion
  in the common-envelope context have successfully employed 2D simulations as a 
  first-order approximation, with subsequent 3D investigations confirming the main qualitative 
  trends while revealing some quantitative differences \citep{Gracia-Linares2015,Kim2024}.

%
\section{Results}
\label{sec:Results}

The morphology of the evolved common envelope surrounding a rotating black hole 
is illustrated in Figures \ref{fig:Density} and \ref{fig:Mach}, 
which depict representative cases by systematically varying the black hole spin, 
Newtonian Mach number, and density gradient. The steady-state accretion rates and
Bremsstrahlung luminosity, measured at approximately 20 dynamical times $t\sim 20\, r_{\rm a}/v_{\infty}$
are summarized in Figure \ref{fig:Rates}, and quantified through analytical 
fitting formulas provided in equations \eqref{eq:fitmrate}--\eqref{eq:fitlrate}. 
To ensure a comprehensive exploration of the parameter space, we examined 
a wide range of configurations: five distinct values of black hole spin, encompassing 
both co-rotating and counter-rotating scenarios, six models of density gradients, and 
six values of the Mach number. This resulted in a total of 66 unique models, 
as detailed in Tables \ref{tab:params1} and \ref{tab:params2}. 
This extensive parameter study allows for a robust analysis of the system's 
behavior under varying physical conditions.

\subsection{Flow Morphology}

\begin{figure*}
\begin{center}
\includegraphics[width=0.95\columnwidth]{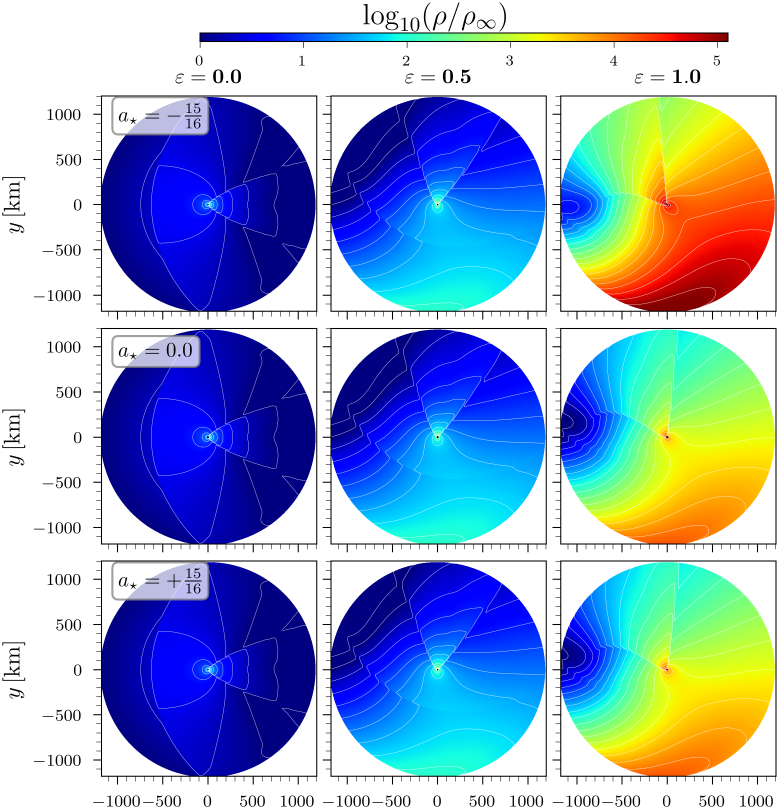} \hspace{-0.15cm}
\includegraphics[width=0.95\columnwidth]{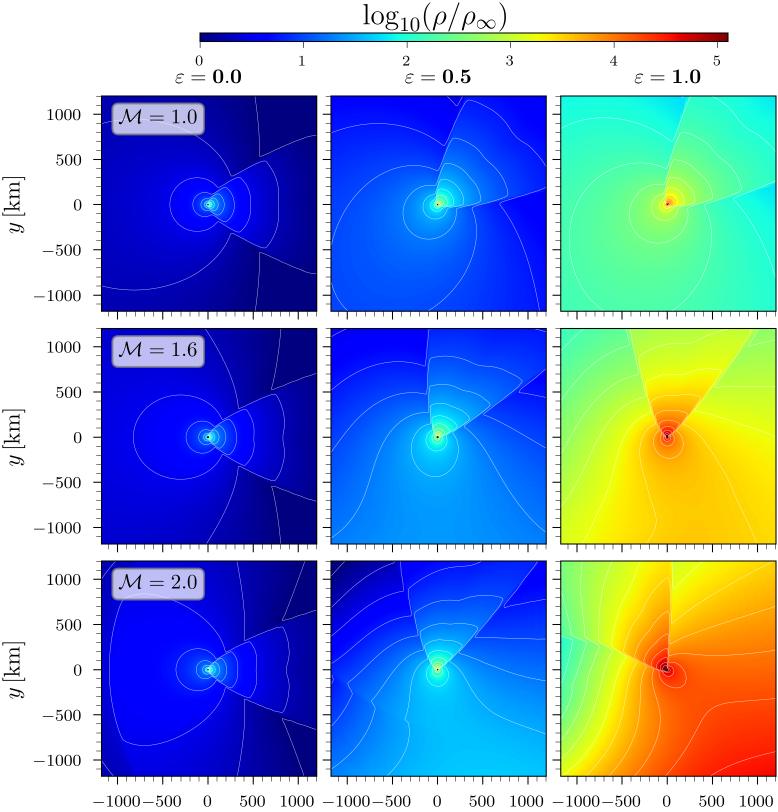}\\
\includegraphics[width=0.95\columnwidth]{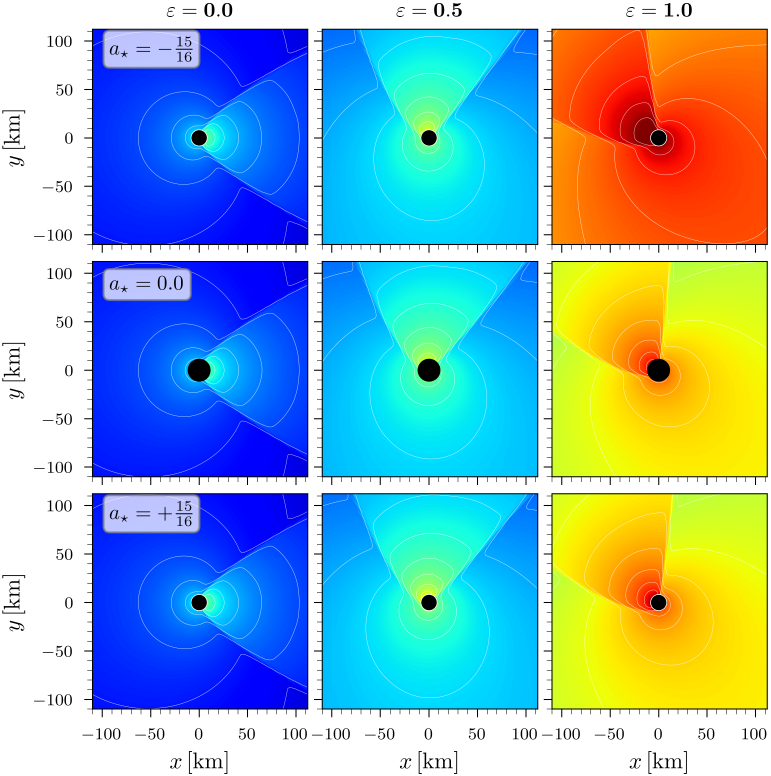} \hspace{-0.15cm}
\includegraphics[width=0.95\columnwidth]{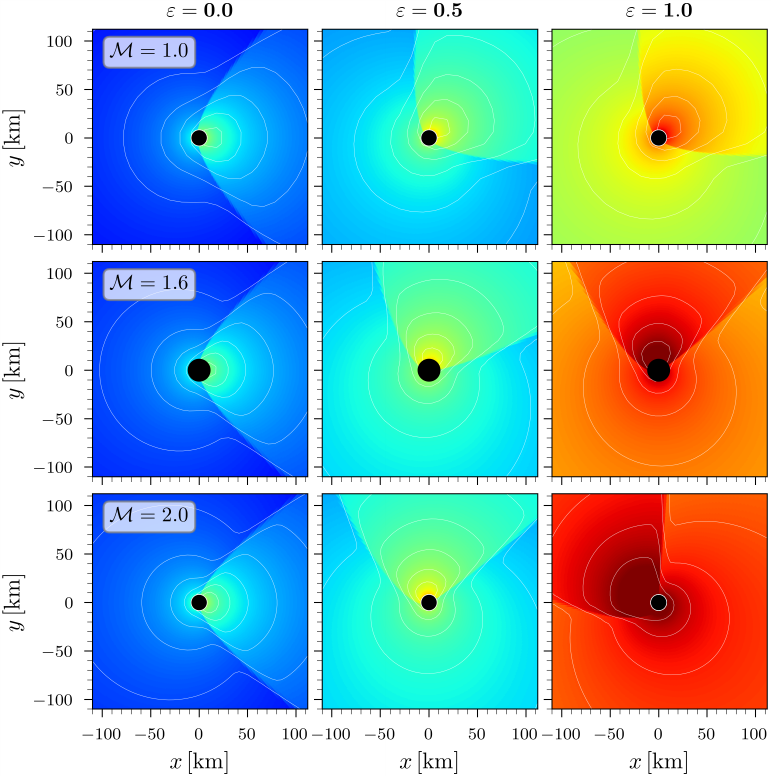}
\caption{Logarithm of the normalised rest mass density.  {\it Left panels} illustrate common envelope 
morphology, when varying the black hole spin $a_{\star}$, and the initial density gradient  $\epsilon_{\rho}$ 
by keeping the Mach number fixed at ${\cal M}_{\infty}= 2.0$, we show only the representative cases. 
The top panels depict the entire numerical domain, while the bottom panels provide a close-up view 
near the black hole vicinity.  {\it Right panels} show the effects on the flow morphology of the Mach 
number and the density gradients for a Schwarzschild black hole. In all cases we show the morphology 
when the flow reaches the steady state, $t \geq 20\, r_{\rm acc}/v_{\infty}$. The white lines depict the 
contours of the normalised rest mass density $ \log_{10} (\rho/\rho_{\infty}) = 0.0,\, 0.25,\, 0.50,\,  ...,\, 1.0.$
}
\label{fig:Density}
\end{center}
\end{figure*}

We begin by analyzing the effects of three key factors on the morphology of 
the common envelope flow: the frame dragging induced by the black hole's rotation, 
the plasma density gradients arising from the black hole's motion near 
the supergiant red star, and the Newtonian Mach number. Across all models,
the formation of a downstream shock cone, a hallmark of Bondi-Hoyle-Lyttleton 
accretion for sonic to supersonic fluids, is observed, consistent with previous studies
\citep{Font98a,Font98c,Font1999b,Donmez2010,Cruz2012,Lora-Clavijo2013,Cruz2013,Cruz2016}. \\

\begin{figure*}
\begin{center}
\includegraphics[width=0.95\columnwidth]{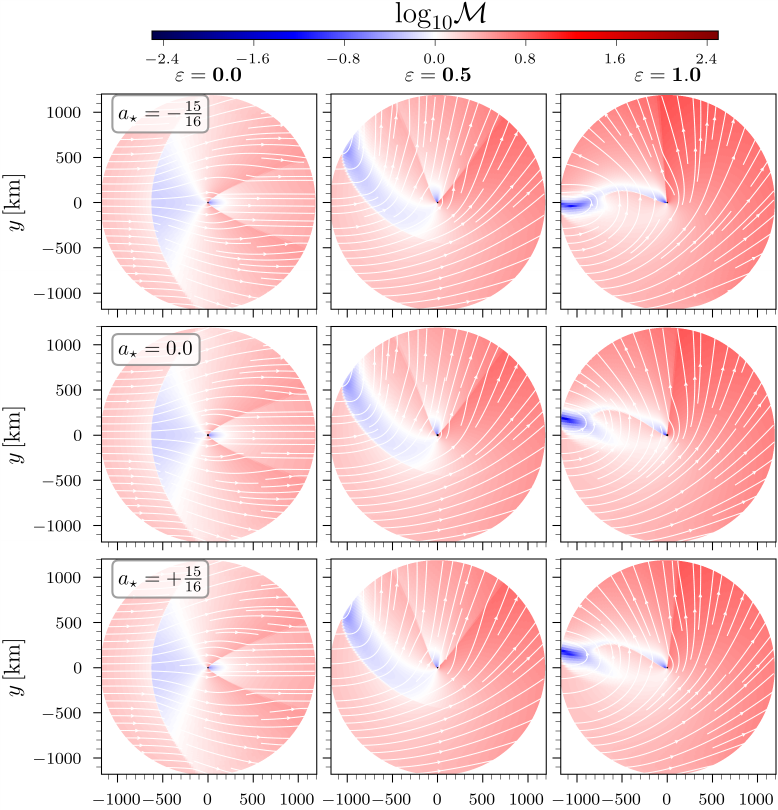} \hspace{-0.15cm}
\includegraphics[width=0.95\columnwidth]{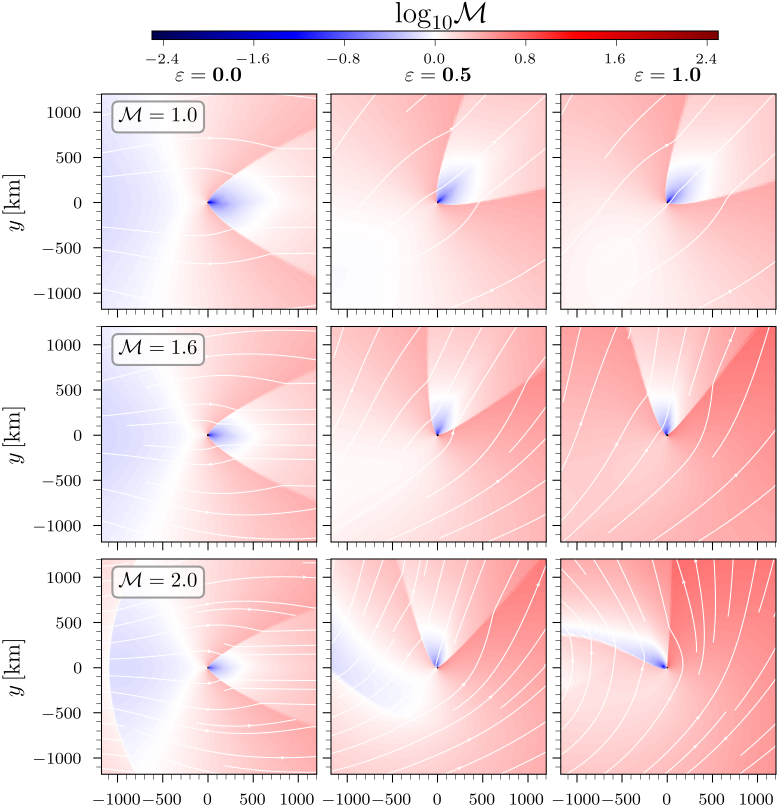}\\
\includegraphics[width=0.95\columnwidth]{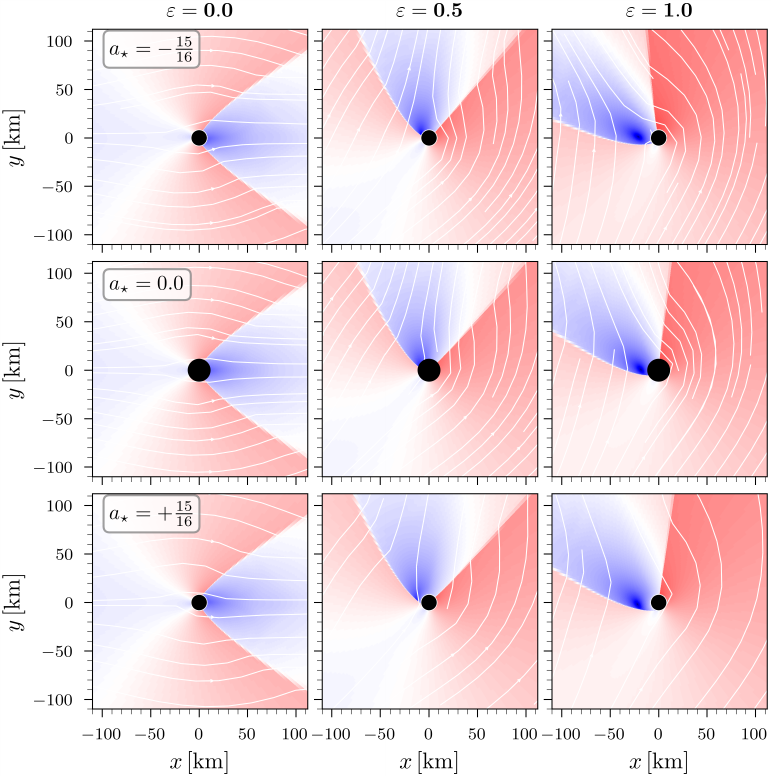} \hspace{-0.15cm}
\includegraphics[width=0.95\columnwidth]{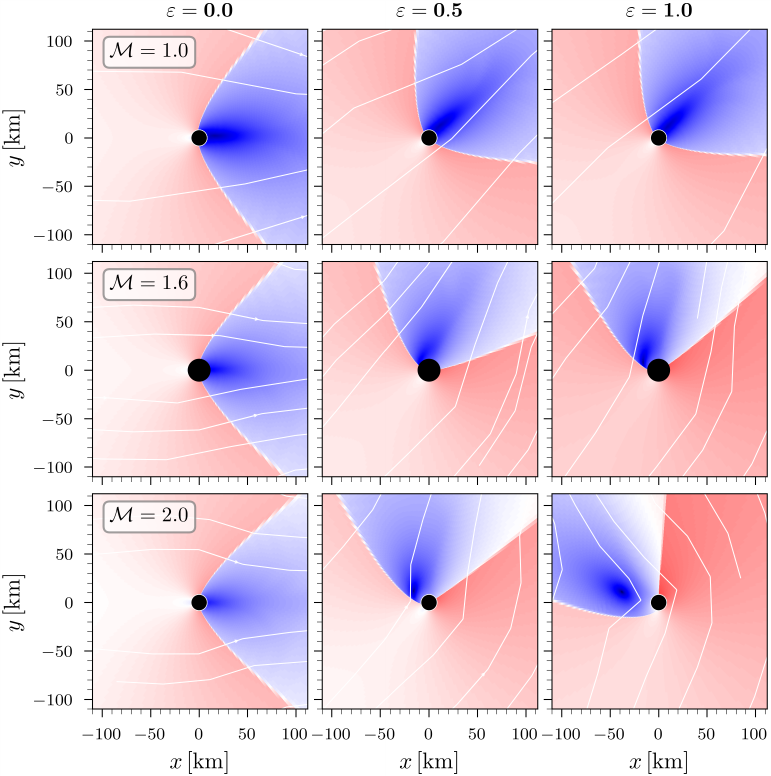}
\caption{Logarithm of the Mach number of the common envelope. Similarly to figure \ref{fig:Density}, 
{\it left panels} shown representative models for black hole spin versus density gradient with fixed Mach 
number ${\cal M}_{\infty}= 2.0$. The {\it Right panels} illustrate models for Newtonian Mach number versus density 
gradients  for a Schwarzschild black hole. Streamlines of the fluid velocity are represented by 
white lines. Subsonic regions are shown in blue, while supersonic flows are indicated in red.}
\label{fig:Mach}
\end{center}
\end{figure*}

In Figures~\ref{fig:Density} and~\ref{fig:Mach}, we present the steady-state gas 
morphology for representative values of the common-envelope density gradient 
and Mach number. We focus on the rest-mass density and Mach number distributions 
to characterize the structure of the shocked flow and the transitions between 
subsonic and supersonic regimes.\\
By fixing the asymptotic Mach number to $\mathcal{M}_{\infty}=2$ and systematically 
varying the black hole spin -- counter-rotating, non-rotating, and co-rotating configurations 
and the density gradient, $0 \leq \varepsilon \leq 1$ -- corresponding to progressively 
deeper regions within the common envelope, two main morphological effects emerge 
in both the density and Mach number distributions (left panels of Figures~\ref{fig:Density} 
and~\ref{fig:Mach}). The first is frame dragging in the vicinity of the black hole, while the 
second is produced from the pressure gradient induced by the background density 
stratification. Together, these effects generate a net force perpendicular to the orbital 
motion, causing the shock cone to be deflected toward regions of lower pressure, i.e., 
away from the core of the red supergiant. The deflection angle increases systematically 
with increasing density gradient for all models with $\varepsilon>0$.\\
On larger scales, a bow shock develops upstream of the black hole. This structure 
produces a clear density discontinuity that separates the incoming supersonic flow 
from the shocked subsonic gas (see bottom panels). In the absence of a density 
gradient, the bow shock remains symmetric with respect to the $x$-axis. As the 
gradient increases, however, the bow shock becomes progressively displaced toward 
lower-density regions. For $\varepsilon=0.5$, the asymmetry is already pronounced, 
while for the largest gradients the bow shock merges with the shock cone. The 
interaction between the shocked gas and the incoming envelope material reduces 
the flow velocity, producing an extended subsonic region downstream of the bow 
shock, shown in blue in Figure~\ref{fig:Mach}. For the largest density gradients, localized 
reverse flows also appear in the downstream region.\\
Near the black hole, the rest-mass density increases by up to four orders of magnitude 
as the density gradient is increased. Simultaneously, the flow undergoes a transition 
from supersonic ($\mathcal{M}>1$) to subsonic ($\mathcal{M}<1$) regimen as it 
crosses the shock cone. The gas reaching the black hole therefore originates 
predominantly from the shocked, transonic or subsonic region downstream of the 
bow shock. As the shock cone becomes increasingly deflected, a stagnation-point-like 
structure develops asymmetrically on one side of the flow (see blue regions inside the 
shock in bottom panels of Fig.~\ref{fig:Mach}). Although the black hole spin has only 
a modest impact on the large-scale morphology, frame-dragging effects are more 
visible close to the event horizon, where they introduce small but measurable 
distortions in the flow structure.\\

 In the second set of simulations, we investigate the morphology of the common-
 envelope flow by varying the asymptotic Mach number and the density gradient, 
 fixing the black holes spin to zero. 
 The right panels of Figures~\ref{fig:Density} and~\ref{fig:Mach} show the 
 steady-state gas morphology for representative values of the density gradient 
 ($\varepsilon = 0.0, 0.5, 1.0$) and Mach number ($\mathcal{M}_{\infty} = 1.0, 1.6, 2.0$).
In addition to the shock-cone deflection induced by the density gradient discussed 
above, both the density and Mach number morphology reveal a systematic 
displacement of the bow shock away from the black hole as the flow transitions 
from supersonic to transonic regimes. These results confirm the expected dependence 
of the shock cone opening angle on the Mach number, in agreement with previous 
studies \citep{Lora-Clavijo2015}.
This behaviour is particularly evident in the inset panels, where the shock cone exhibits 
a large opening angle for the transonic case, $\mathcal{M}{\infty}=1.0$, which 
progressively decreases as the Mach number increases to $\mathcal{M}{\infty}=2.0$. 
We also find that frame-dragging effects become less pronounced in transonic flows 
than in supersonic ones.
As in the previous set of simulations, models without a density gradient exhibit a 
symmetric flow morphology and a stagnation-point-like structure located symmetrically 
within the shock cone. As the density gradient increases, the shock cone is 
progressively deflected toward lower-density regions, and the stagnation-point-like 
structure becomes increasingly asymmetric.\\
In summary, by combining the effects of black hole spin, common-envelope 
density gradients, and variations in the Mach number through changes in the 
gas velocity and sound speed, we obtain a rich variety of flow morphologies. 
On large scales, the simulations exhibit bow shocks and downstream shock 
cones consistent with previous studies \citep{Font1999b,Lora-Clavijo2015,Cruz2020b}. 
Near the black hole, additional distortions produced by relativistic frame 
dragging become important. Our results also confirm the dependence of the 
shock-cone opening angle on the degree of flow Mach number, in agreement 
with the findings of \citet{Lora-Clavijo2015}. In the absence of density gradients, 
we recover the symmetric bow-shock and shock-cone morphologies reported 
in our previous work \citep{Cruz2020b}, while increasing density gradients 
introduce progressively stronger asymmetries and shock-cone deflections.
 
\subsection{Accretion Rates}

The second part of our analysis focuses on quantifying how much 
the plasma in the common envelope contributes to the black hole's 
mass and angular momentum. In each simulation, 
we measure the mass accretion rate, $\dot{M}_0$, at the 
event horizon\footnote{$ \dot{M} = \int_{0}^{2\pi} \alpha\sqrt{\gamma}
W \rho (v^{r} - \beta^{r}/\alpha) d\phi$, where $W$ is the Lorentz factor, 
$\alpha$ is the lapse function 
and $\sqrt{\gamma}$ is the determinant of the three metric.}. 
In Equation \eqref{eq:fitmrate}, we summarize the effects of black 
hole spin, density gradients, and Mach number using a fitting 
formula for the normalized mass accretion rate. Here, $\dot{M}_{0,\rm ref}$
 is the reference rate from Bondi accretion\footnote{ $\dot{M}_{0,\rm ref}= 4 \pi \lambda \rho_{\infty} 
M^{1/2} r^{3/2}_{\rm acc}$, where $\lambda \simeq 0.71$.}. 
The coefficients in the fitting formula 
are $\mu_{i} = (7.82 \times 10^{-3}, 1.3, 2.76, 2.94)$. 

\begin{figure*}
\begin{center}
\includegraphics[width=1.85\columnwidth]{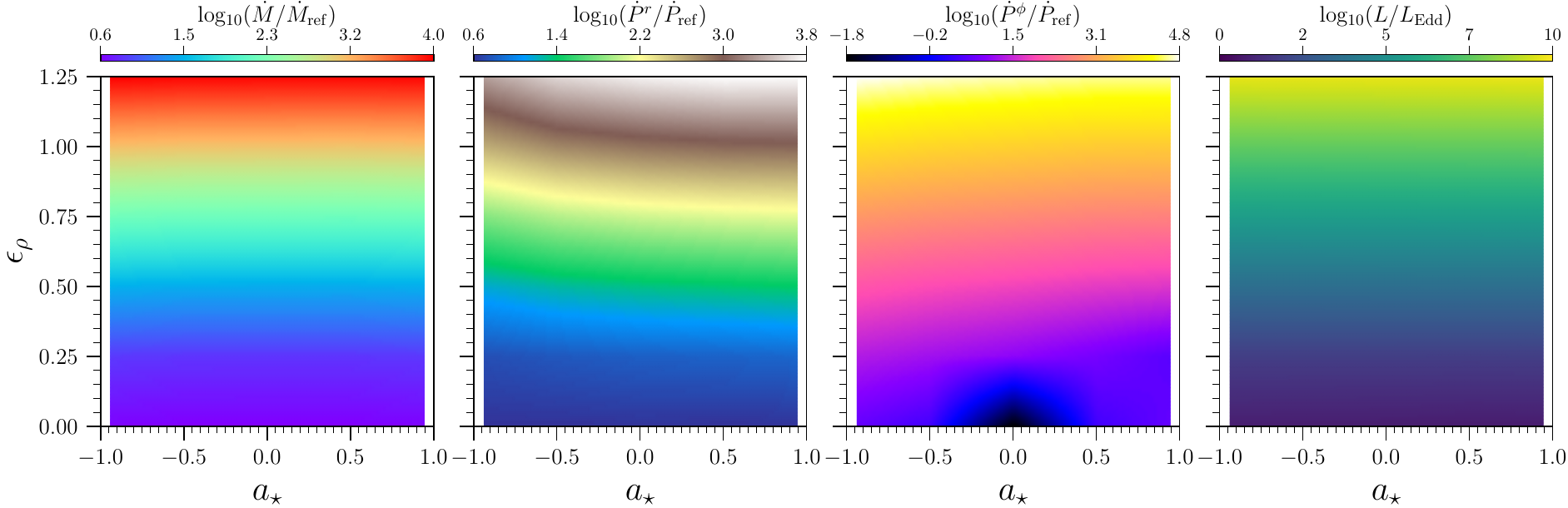}\\
\includegraphics[width=1.85\columnwidth]{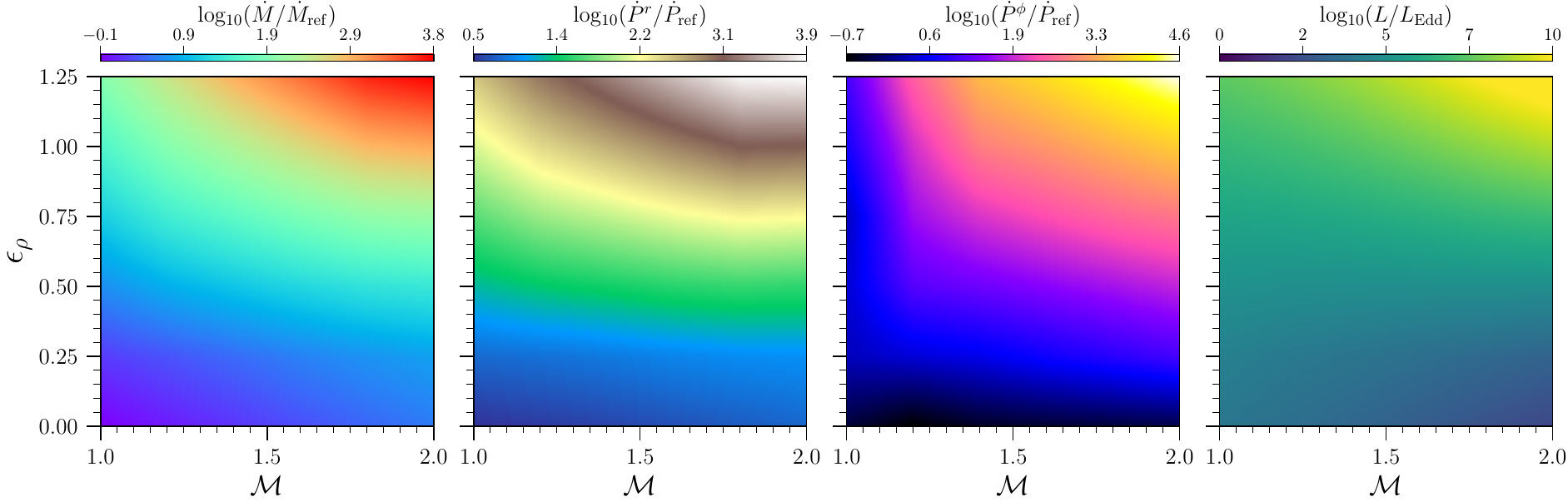}
\caption{Mass accretion rates, radial and angular momentum, and bremsstrahlung luminosity. 
In the {\rm top panels} we show the two dimensional accretion rates as function of the black hole 
spin $a_{\star}$ and density gradient $\epsilon_{\rho}$.  In the {\rm Bottom panels} we show the accretion rates 
as function of the Newtonian Mach number $\cal M$ and the density gradient $\epsilon_{\rho}$.}
\label{fig:Rates}
\end{center}
\end{figure*}

The left panels of Figure \ref{fig:Rates} show the normalized 
mass accretion rate, with the top panel displaying black hole 
spin versus density gradients and the bottom panel showing 
Mach number versus density gradients. Our results reveal that 
the density gradient has a more significant impact on the 
mass accretion rate compared to black hole spin and Mach number. 
Specifically, we observe that as the stellar black hole moves 
closer to the supergiant red star and travels at higher 
Mach numbers, the amount of fluid accreted from the 
common envelope increases significantly.

The rates of radial and angular momentum\footnote{$\dot{P}^{i}=-\int_{\partial V} 
\alpha \sqrt{\gamma} T^{i j} \, d\Sigma_{j} + \int_{V} \alpha \sqrt{\gamma} S^{i} \, dV$, 
where $T^{i j}$ are the spatial components of the energy-momentum tensor, 
and $S^i$ is the momentum source term.}, $\dot{P}^{\phi}$ and $\dot{P}^{r}$, 
are described by fitting formulas presented in 
Equations \eqref{eq:fitprrate} and \eqref{eq:fitpprate}. 
These formulas are expressed in terms of the momentum rate for 
Bondi accretion\footnote{$\dot{P}_{\rm ref} = 
\dot{M}_{0,\rm ref} \frac{v_{\infty}}{\sqrt{1-v^{2}_{\infty}/c^{2}}}$}, $\dot{P}_{\rm ref}$. 
The coefficients for the radial and angular momentum 
components are $\mathscr{r}_{i}  = (0.064,\ -0.045,\ 1.240,\ 2.44,\ 2.93,\ 0.94,\ -0.20)$ 
and $\mathscr{p}_{i} = (-0.054,\ 0.021,\ -0.087,\ 8.66,\ -0.19,\ 0.30,\ 0.12)$, respectively.

The behavior of the momentum rates, determined by combining the 
degrees of freedom parameters, is illustrated in the second and third 
rows of Figure \ref{fig:Rates}. From these results, we observe that 
the transfer of radial momentum from the common envelope 
to the black hole increases with both the density gradient and 
the Mach number, mirroring the trend seen in mass accretion. 
In contrast, the angular momentum exhibits a minimum accretion 
rate for a non-rotating black hole and small Mach number.

\begin{eqnarray}
  \log\left(\frac{\dot{M}_0}{\dot{M}_{0,\rm ref}}\right)&=&\left( 1 + \mu_{0}a_{\star}^2 \right) \left(  \mu_{1} + \mu_{2}\epsilon_{\rho} + \mu_{3}\epsilon^{2}_{\rho} \right) \frac{\cal M}{2}\,,   \label{eq:fitmrate} \\
  \log \left(\frac{\dot{P}^{\phi}}{\dot{P}_{\rm ref}}\right) &=&\left( 1 + \mathscr{p}_{0}a_{\star} +\mathscr{p}_{1}a_{\star}^2 \right)  \left(  \mathscr{p}_{2} +\mathscr{p}_{3}\epsilon_{\rho} + \mathscr{p}_{4}\epsilon^{2}_{\rho} \right)\times \nonumber\\
  &&\left( \mathscr{p}_{5}{\cal M}+\mathscr{p}_{6}{\cal M}^2\right), \label{eq:fitprrate}\\ 
    \log \left(\frac{\dot{P}^{r}}{\dot{P}_{\rm ref}}\right) &=&\left( 1 +\mathscr{r}_{0}a_{\star}+  \mathscr{r}_{1}a_{\star}^2 \right) \left(  \mathscr{r}_{2} +\mathscr{r}_{3}\epsilon_{\rho} + \mathscr{r}_{4}\epsilon^{2}_{\rho} \right)\times \nonumber\\
  &&\left( \mathscr{r}_{5}{\cal M}+\mathscr{r}_{6}{\cal M}^2\right), \label{eq:fitpprate} \\
\end{eqnarray}
\begin{eqnarray}
    \log_{10} \left(\frac{L_{_{\rm BR}}}{L_{\rm Edd}}\right) &=& \left( 1 +\ell_{0} a_{\star} \right)(\ell_{1}+\ell_{2}\epsilon_{\rho} + \ell_{3}\epsilon^{2}_{\rho}) \times \nonumber\\
    &&( \ell_{4}+  \ell_{5}{\cal M}^{3/2} ) \,.    \label{eq:fitlrate}
\end{eqnarray}

\subsection{Bremsstrahlung luminosity}

We compute the bremsstrahlung luminosity arising from electron-proton collisions\footnote{
$L_{_{\rm BR}}= 3.0 \times 10^{78}  \int \left( W \, T^{\frac{1}{2}} 
\rho^{2} \sqrt{\gamma} dV \right) 
\left( \frac{M_{\odot}}{M} \right) \ {\rm  \frac{erg}{s}}$, where $T=p/\rho$ 
is the dimensionless temperature.}  as a diagnostic of the local thermal state 
of the shocked plasma. In the dense, optically thick environment of a common 
envelope, photons produced near the black hole do not escape freely; they are 
trapped and reprocessed via diffusion before emerging from the stellar surface. 
Consequently, the bremsstrahlung luminosity computed here does not represent 
observable radiation. Instead, it serves as a comparative measure across our 
parameter space ($a_{\star}$, $\mathcal{M}$, $\varepsilon$) to assess how 
black hole spin and flow conditions affect the post-shock temperature and 
density structure. For typical envelope parameters, 
the bremsstrahlung cooling timescale 
is orders of magnitude longer than the dynamical timescale 
\citep{Rybicki_Lightman1986}, 
indicating that radiative cooling
is not dynamically important in the present simulations. We therefore 
emphasize that the bremsstrahlung luminosity is a diagnostic tool rather
than a prediction of observable emission. 
The result is quantified in Equation \eqref{eq:fitlrate}. 
For convenience, and to allow rescaling for arbitrary black hole masses, 
the luminosity is normalized by the Eddington luminosity, 
$L_{\rm Edd} = 1.26 \times 10^{38} \, (M/M_{\odot}) \, {\rm erg\,s^{-1}}$. 
The coefficients in the fitting formula are given by $\ell_{i} = (0.0021,\ 0.577,\ 5.318,\ 1.655,\ 3.0,\ 3.1)$.
The right panels of Figure \ref{fig:Rates} illustrate the two-dimensional 
morphology of luminosity as 
a function of the black hole spin, Mach number, and density gradients. 
We observe a behavior similar
to that of the mass accretion rate, which supports the assumptions 
in the literature that these two 
quantities are proportional, \ie $\dot{M}_0 \propto L{_{\rm BR}}$.

\section{Conclusions}
\label{sec:conclusion}

This paper investigates the morphology, accretion dynamics, and bremsstrahlung luminosity of a 
supergiant red star common envelope interacting with a stellar rotating black hole under varying 
conditions of spin, Mach number, and density gradients, ($a_{\star},\, {\cal M},\  \varepsilon$). 
The simulations reveal that the formation of a downstream shock cone is a universal feature of 
Bondi-Hoyle-Lyttleton accretion in sonic to supersonic regimes. Black hole rotation and density 
gradients induce frame dragging and pressure gradients, resulting in a dragged shock cone whose 
angle increases with larger density gradients. For models with zero density gradients, the shock 
cone remains largely symmetric with minimal dragging; however, higher Mach numbers and stronger 
gradients cause the cone to narrow and shift farther from the black hole. The inclusion of density 
gradients also leads to the emergence of a second bow shock, with the flow between the shocks 
remaining subsonic regardless of black hole spin or Mach number.

The analysis of accretion dynamics reveals that the mass accretion rate significantly increases 
with higher density gradients and Mach numbers -- as the stellar black hole approached to the 
red giant star core--, indicating that the common envelope plasma feeds the black hole more 
efficiently under such conditions. The transfer of radial momentum follows a similar trend, while 
angular momentum accretion is minimal for non-rotating black holes and low Mach numbers. 
Analytical fitting formulas provide a quantitative description of the mass and momentum accretion 
rates, facilitating predictions for broader parameter ranges. Additionally, the study computes 
bremsstrahlung luminosity, normalized by the Eddington luminosity, which demonstrates a behavior 
proportional to the mass accretion rate. This supports the assumption that electromagnetic 
emissions are closely tied to the accretion process. The luminosity morphology highlights the 
significant influence of black hole spin, density gradients, and Mach number on the radiative 
properties of the common envelope. 

Some caveats of our numerical approach arise from the two-dimensional slab geometry 
adopted in this work. In particular, out-of-plane density and pressure gradients are neglected, 
which may introduce biases in volume-integrated quantities such as the mass and momentum 
accretion rates and the bremsstrahlung luminosity. Consequently, these first general relativistic 
2D simulations cannot be used for direct comparisons with common envelope observations, 
and fully three-dimensional studies are required for a robust theoretical--observational comparison. 
Nevertheless, our simulations capture the essential physical behaviour of a rotating black hole 
interacting with a common envelope and provide first insight into the nonlinear dynamics of the 
common envelope phase. Future work will extend this analysis to three dimensions, enabling 
more realistic theoretical-observational comparisons.

\section*{Acknowledgements}
The authors would like to thank Luciano Rezzolla for
helpful discussions related to this work. 
This research was supported by DGAPA-UNAM (grant IN110522) and the
Ciencia B\'asica y de Frontera 2023-2024 program of SECIHTI M\'exico
(projects CBF2023-2024-1102 and 257435). 
F.D.L-C is supported by the Vicerrectoría de Investigación y Extensión - 
Universidad Industrial de Santander, under Grant No. 3703.
The simulations were performed on Going Merry workstation, Atocatl-LAMOD 
cluster at UNAM, and on the  Calea cluster at the ITP in Frankfurt.

\section*{Data Availability}
The data underlying this article will be shared on reasonable request
to the corresponding author.



\appendix

\bsp	
\label{lastpage}
\end{document}